\documentclass[%
 preprint,
 amsmath,amssymb,
 aps,
prb,
]{revtex4-2}

\usepackage{graphicx}
\usepackage{dcolumn}
\usepackage{bm}
\usepackage{enumitem}
\usepackage[dvipsnames]{xcolor}

\begin{document}


\title{Exploring x-ray irradiation conditions for triggering ultrafast diamond graphitization}

\author{Vladimir Lipp}
\email{vladimir.lipp@desy.de}
\affiliation{Center for Free-Electron Laser Science CFEL, Deutsches Elektronen-Synchrotron DESY, Notkestr. 85, 22607 Hamburg, Germany}

\author{Victor Tkachenko}
\affiliation{European XFEL, Holzkoppel 4, 22869 Schenefeld}
\affiliation{Center for Free-Electron Laser Science CFEL, Deutsches Elektronen-Synchrotron DESY, Notkestr. 85, 22607 Hamburg, Germany}

\author{Ichiro Inoue}
\affiliation{RIKEN SPring-8 Center, 1-1-1 Kouto, Sayo 679-5148, Hyogo, Japan}

\author{Philip Heimann}
\affiliation{Linac Coherent Light Source, SLAC National Accelerator Laboratory, 2575 Sand Hill Road, Menlo Park, California 94025, USA}

\author{Anastasiia Ryzhkova}
\affiliation{Taras Shevchenko National University of Kyiv, 64/13, Volodymyrska Street, 01601 City of Kyiv, Ukraine}

\author{Abdelkhalek Bashandi}
\affiliation{Department of Physics, Faculty of Science, Cairo University, 1 Gamaa Street, 12613 Giza, Egypt}

\author{Beata Ziaja}
\email{beata.ziaja-motyka@cfel.de}
\affiliation{Center for Free-Electron Laser Science CFEL, Deutsches Elektronen-Synchrotron DESY, Notkestr. 85, 22607 Hamburg, Germany}
\affiliation{Institute of Nuclear Physics, Polish Academy of Sciences, Radzikowskiego 152, 31-342 Krak\'ow, Poland}

\date{\today}

\begin{abstract}
Intense femtosecond x-ray pulses produced by an x-ray free-electron laser can trigger irreversible structural transitions in crystalline solids. For instance, irradiation of diamond can lead to graphitization and, at higher deposited doses, to amorphization. Our Monte Carlo simulations of irradiated diamond under realistic experimental conditions demonstrate that triggering graphitization or other phase transitions with hard x-ray photons can be challenging due to the ballistic escape of photoelectrons out of the beam focus. Decisive parameter here is the photoelectron range in proportion to the focal beam size. For future experiments on x-ray-induced transitions, such dedicated simulations of ballistic transport preceding the beamtime will be necessary. They can predict experimental conditions under which the desired distribution of the absorbed x-ray dose in the irradiated solid can be achieved.
\end{abstract}

\maketitle


\section{\label{sec:intro}Introduction}

Femtosecond x-ray pulses produced by the modern x-ray free-electron lasers (XFELs) \cite{FLASH,FERMI,LCLS,ishikawa2012compact,weise2017commissioning} can be sufficiently intense to cause  irreversible structural damage to crystalline solids, including radiation-hard diamond \cite{Gaudin2013Photon,Inoue2016Observation,Heimann2023Nonthermal}.

Within first hundred femtoseconds after an absorption of an x-ray pulse, the electron subsystem in an irradiated solid undergoes a strongly nonequilibrium evolution triggered by the so-called electron cascading \cite{Ziaja2001cascades,Follath2019B4C,medvedev2015femtosecond}. In case of hard x-rays, energetic photo-, Auger and secondary electrons inelastically scatter on the target atoms, creating an avalanche of excited secondary electrons. On the ultrafast timescales considered, the excited electrons propagate ballistically \cite{lipp2022quantifying}, in contrast to a slow diffusive motion of low-energy conduction-band electrons. The electron cascading finishes when all the cascade electrons lose most of their energy on impact ionizations and can no longer excite secondary electrons. The resulting distribution of low-energy excited electrons in the conduction band may affect the atomic potential energy surface and lead to a nonthermal transition, such as disordering \cite{Heimann2023Characterization,Heimann2023Nonthermal} or graphitization \cite{Tavella2017graphitization}, on  femtosecond timescales. This occurs when a sufficient number of electrons is excited in the conduction band \cite{silvestrelli1996ab,kudryashov2001band,rousse2001non,harb2008electronically,XTANT2013}.

According to experimental and theoretical studies in refs. \cite{Tavella2017graphitization,Gaudin2013Photon},  diamond irradiated with extreme ultraviolet (XUV) pulse undergoes graphitization on a timescale of about 150 fs, if the absorbed dose reaches about 1 eV/atom (see also the update of the dose value in ref. \cite{lipp2022density}). It is to be emphasized that the final state of the material after this time is an overdense graphite \cite{medvedev2013nonthermal}, i.e., only sp3 $\rightarrow$ sp2 rebonding takes place during 150 fs. Another simulation performed with a hybrid Monte Carlo/Density Functional Tight Binding/Molecular Dynamics tool XTANT+ indicated that the graphitization threshold is about 2 eV/atom and only a partial graphitization can be expected on a timescale of about 500 fs \cite{lipp2022density}. 

However, in a more recent work by Heimann et al. \cite{Heimann2023Nonthermal}, the attempt to directly observe graphitization on femtosecond time scales, induced by hard x rays, was unsuccessful. Diamond disordering was observed instead. Also in the recent work by L\"utgert et al. \cite{Lutgert2024diamond}, the authors found no evidence of the diamond graphitization caused by the heavy ion beam at a similar dose deposition.

In this work, we perform dedicated Monte Carlo simulations with the XCASCADE-3D code, in order to investigate the effect of ballistic electron transport on the  hard x-ray-induced ultrafast structural transformations in diamond. The irradiation parameters are taken from the recent experimental work \cite{Heimann2023Nonthermal}. We show that the ballistic electron motion can result in a strongly non-uniform spatial redistribution of the transient number of low-energy conduction-band electrons  in the irradiated crystal. According to earlier studies, a sufficient number of such electrons is needed to initiate the graphitization process \cite{medvedev4open,Tavella2017graphitization,lipp2022density}. The escape of the electrons out of the beam focus can yield  effective absorbed x-ray doses much lower than those expected after assuming a full absorption of the pulse within the lateral beam focus. This shows that dedicated simulations, preceding planned graphitization and other phase-transition experiments induced by x rays, may be necessary in order to establish  proper irradiation conditions needed for such experiments.

\section{\label{sec:exp}Experimental parameters}

Parameters of the present simulations correspond to the recently published experimental study on the nonthermal structural transformation in diamond performed at the SACLA XFEL facility \cite{Heimann2023Nonthermal}. In that study, a nanocrystalline diamond layer with the thickness of $L=20$~{\textmu}m was irradiated with a 6-fs x-ray pulse. XFEL photon energy was 7 keV. At SACLA, the beam profiles can be well described by a Gaussian function \cite{yumoto2013focusing,mimura2014generation,yumoto2020nanofocusing,yamada2024extreme}. Various pulse energies $E_{beam}$, spot areas $S$,  and, consequently, different volumetric nominal absorbed doses $D$, were applied. The dose is defined as:


\begin{equation}
D = E_{beam} \times (1-\exp(-L/\lambda_{att})) \frac{1}{n_a L\times S} ,
\end{equation}


\noindent where $n_{a}=176$~nm$^{-3}$ is the average atomic number density for diamond under ambient conditions.

In the experiment \cite{Heimann2023Nonthermal}, the diamond x-ray pump-probe analysis included several data sets. They were combined into three groups as shown in Table \ref{tab:irradiation_parameters}. The foci are defined as full width at half maximum (FWHM) spots.

\begin{table}[h]
	\centering
	\begin{tabular}{|c|c|c|c|c|}
		\hline
		Experimental group & D (eV/atom) & Focus ({\textmu}m$^2$)                  & $E_{beam}$ ({\textmu}J) & Calculated peak electron density (\%) \\
		\hline
		\hline
		1.1              & 1.3         & 1.83 $\times$ 1.15             & 33          & 1.6 \\
		\cline{1-5}
		1.2              & 1.4         & 1.44 $\times$ 1.10             & 27          & 1.7 \\
		\cline{1-5}
		1.3              & 1.5         & 1.06 $\times$ 1.05             & 21          & 1.8 \\
		\cline{1-5}
		1.4              & 1.5         & 0.92 $\times$ 0.86             & 15          & 1.7 \\
		\hline
		\hline
		2.1              & 5.2         & 0.78 $\times$ 0.66             & 33          & 5.5 \\
		\cline{1-5}
		2.2              & 5.7         & 0.59 $\times$ 0.51             & 21          & 5.0 \\
		\cline{1-5}
		2.3              & 8.7         & 0.40 $\times$ 0.35             & 15          & 5.9 \\
		\hline
		\hline
		3.1              & 16          & 0.40 $\times$ 0.35             & 27          & 10  \\
		\cline{1-5}
		3.2              & 54          & 0.14 $\times$ 0.16             & 15          & 12  \\
		\hline
	\end{tabular}
	\caption{Irradiation parameters used in Ref. \cite{Heimann2023Nonthermal} combined into three data sets.}
	\label{tab:irradiation_parameters}
\end{table}
\noindent
\section{\label{sec:model}Model}

Simulations presented in this work have been performed using our in-house classical Monte Carlo code XCASCADE-3D \cite{XCASCADE3D}, which was first introduced in ref.  \cite{lipp2017MC}. Later, it was used to explain a number of experimental observations in ultrafast x-ray science, such as damage to ruthenium mirrors induced by XUV light and hard x-rays \cite{milovlipp}, or transient optical properties of irradiated Si$_3$N$_4$ \cite{TkachenkoLipp2021}. The tool also enabled us to identify suitable materials or pulse parameters  in various x-ray-related applications \cite{lipp2022quantifying}.

The XCASCADE-3D simulation tool can provide temporal and spatial characteristics of x-ray-induced electron cascades in various solids on femtosecond time scales. In particular, the code resolves fast electron ballistic transport, which affects the distribution of the x-ray dose absorbed. In order to describe photoabsorption and electron scattering on atoms, the code uses an atomistic approximation: it applies atomic cross sections (for isolated atoms and ions) and atomic ionization potentials. They are extracted from the EPICS 2017 database \cite{cullen2018survey}. Decay of the created deep-shell holes is simulated by taking into account the characteristic Auger decay times from ref. \cite{keski1974total}. More details can be found in ref. \cite{lipp2022quantifying}. 

The XCASCADE-3D results, presented here, were averaged over up to 1 million Monte-Carlo realizations to achieve reliable statistics. The grid size for the figures was 10~nm $\times$ 10~nm $\times$ 10~nm. The beam propagation axis was chosen to be parallel to Z axis (directed in-depth into the material) both in the simulations and in the plots.  X-ray polarization direction was linear and parallel to the X axis. In the present simulations, similarly to Ref. \cite{milovlipp}, it was taken into account by assuming that a half of the photoelectrons started their motion in the direction +X, and the other half did in the direction -X. In reality, a cosine angular distribution is expected \cite{DavisClassical2022} (eq. (3.29) therein), i.e., we somewhat overestimate the effect of the electron spread with our simplifying assumption. The anisotropic, cosine photoelectron distribution will be implemented into a future version of the software. 
The photoelectrons underwent  elastic and inelastic scatterings, leading to the creation of Auger and secondary electrons, until their energy fell  below the lowest ionization threshold for diamond. According to the EPICS database \cite{cullen2018survey}, the ionization threshold for a carbon atom in diamond is 11.26 eV.  In our simulations, when the energy of a high-energy electron decreases below this value, we assume that the electron stops its motion and becomes an excited electron in the conduction band. To justify this assumption, we estimated the progress of diffusive motion for electrons with the energy of $\sim 11$ eV as $4.35$ A/fs$^{0.5}$, which means that such electrons only travel a distance of $\sim 6$ nm within 200 fs. This distance is much smaller  than, e.g., the thickness of the diamond sample, $20$~{\textmu}m, considered here.

Since the goal of this work is to study possible diamond graphitization caused by low-energy excited electrons in the conduction band, we do not include electrons with energies above 11.26 eV into the plots, assuming that they do not affect the potential energy surface, i.e., they do not trigger graphitization. 

In order to take into account experimental x-ray spot sizes, in this work we implemented a dedicated convolution procedure, which assumes that each electron cascade starts at the site of a photoabsorption and evolves independently. The spatial distributions of the absorbed photons along X and Y axes are Gaussian. The distribution of absorbed x-ray energy along the Z axis is assumed to be homogeneous, due to the large x-ray attenuation length. Namely, for the experimental photon energy, 7 keV, the corresponding photon attenuation length is, $\lambda_{att}=428$~{\textmu}m \cite{henke}. The simulation volume is then a layer of a certain thickness (less than photon attenuation length), infinite in X and Y. Here, it has a thickness of $L=20$~{\textmu}m, as in the experiment \cite{Heimann2023Nonthermal}. 

According to the XCASCADE-3D simulations, the electron cascading time for 7 keV photon in diamond is about 14 fs \cite{lipp2022quantifying}. This means that 14 fs after a photoabsorption event, majority of the excited electrons stop their motion and join the low-energy fraction of conduction-band electrons. The corresponding photoelectron range is about 312 nm \cite{lipp2022quantifying}. 

A typical calculation for each beam parameter takes about 2 hours on Intel(R) Xeon(R) W-2225 CPU @ 4.10GHz, using 8 CPU cores.

\section*{\label{sec:results}Results}

\begin{figure}[h!] 
	
	\centering 
	
	\includegraphics[width=0.45\textwidth]{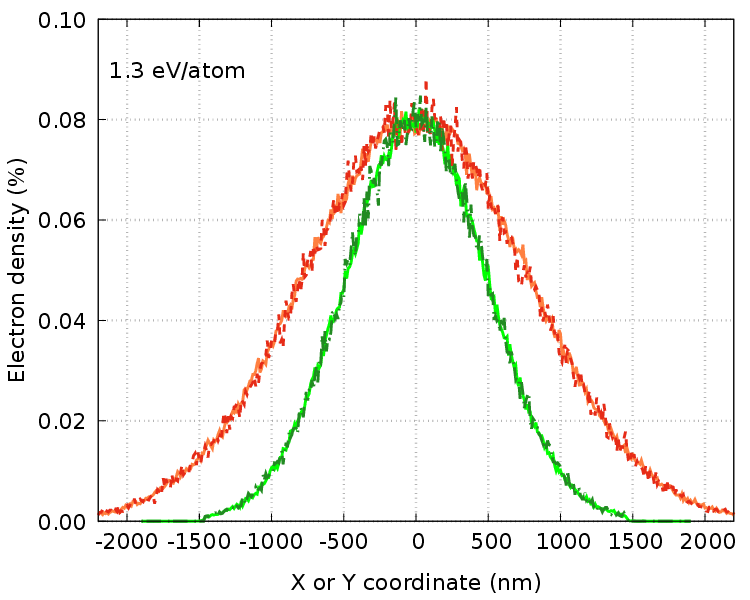} 
	\includegraphics[width=0.45\textwidth]{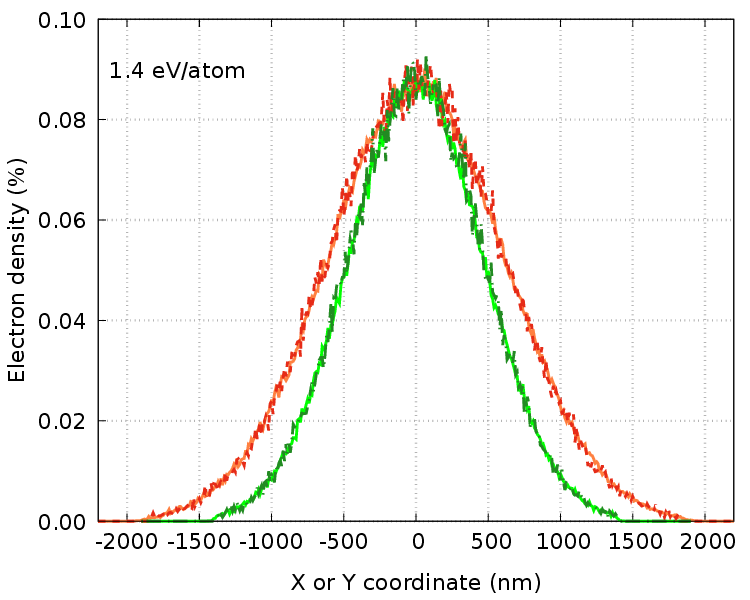} 
	\includegraphics[width=0.45\textwidth]{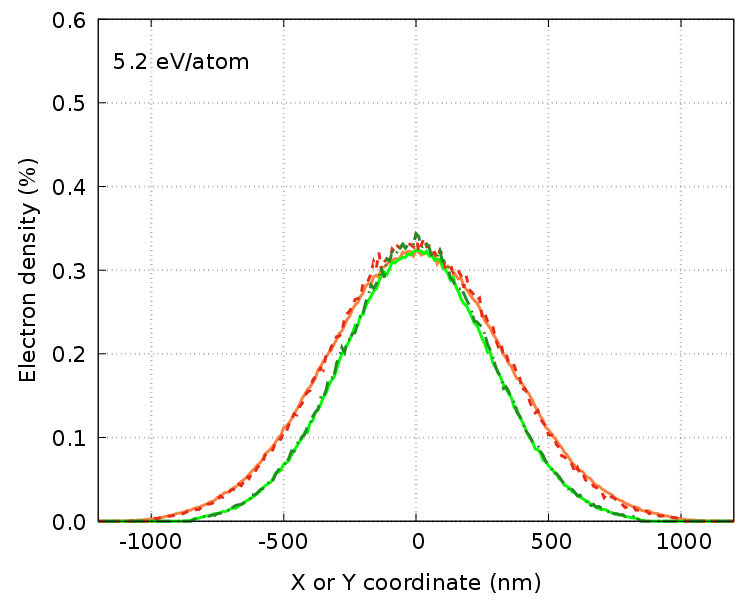} 
	\includegraphics[width=0.45\textwidth]{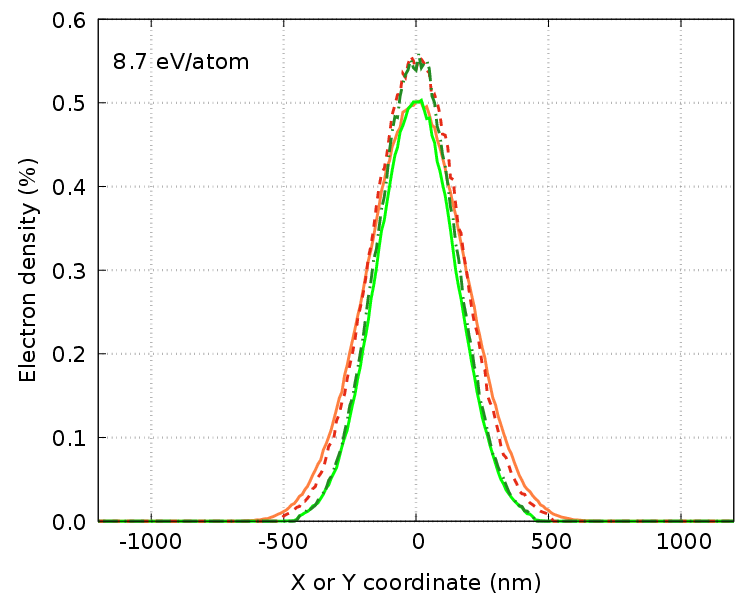} 
	\includegraphics[width=0.45\textwidth]{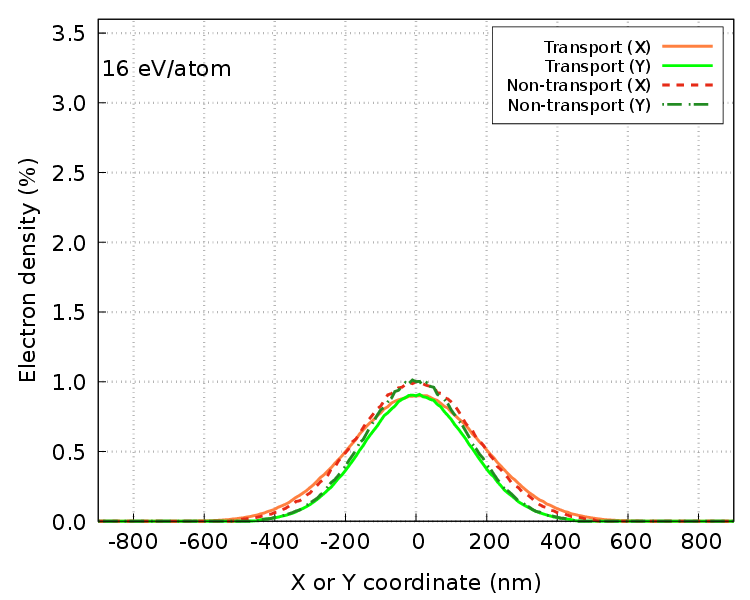} 
	\includegraphics[width=0.45\textwidth]{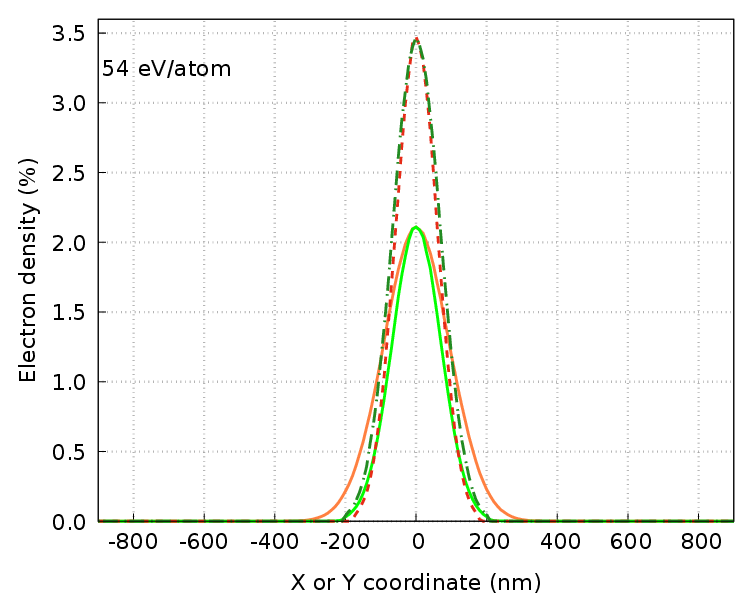} 
	
	\caption{Results of XCASCADE-3D simulations for diamond irradiated with an 6 fs x-ray pulse (photon energy of 7 keV). The solid lines represent the electron density at time $t=0$~fs (i.e., at the temporal maximum of the pulse intensity) for the absorbed doses of: (a) 1.3, (b) 1.4, (c) 5.2, (d) 8.7, (e) 16, (f) 54 eV/atom. Red color corresponds to the direction along the polarization axis X (with $Y=0$), while green color corresponds to the perpendicular direction (with $X=0$). The dashed and dash-dotted lines represent the respective X and Y electron densities estimated in non-transport cases , i.e., they essentially correspond to the X-ray spatial profiles.}
	
	\label{fig:density0fs}

\end{figure}

In Fig.~\ref{fig:density0fs}, we present selected distributions of the electron density in diamond predicted at the temporal maximum of the pulse intensity, $t=0$~fs, with the above-mentioned parameters. Subfigures (a), (b), (c), (d), (e), (f) represent selected experimental cases (1,1), (1,2), (2,1), (2,3), (3,1), and (3,2), respectively. Solid red (green) curves show transient electron density along X (Y) axis, which is parallel (perpendicular) to the beam polarization. Beam propagates along Z axis both in the simulations and in the plots. Dashed (dot-dashed) lines represent the simulated distribution of the absorbed photons (calculated also along X (Y) axis) multiplied by the average number of electrons per a single 7 keV photon in diamond at $t=0$ fs ($\sim 45.6$). In other words, these curves correspond to a simulation not accounting for the electron transport, which effectively means that they reflect the original X-ray spatial profile. We will call these predictions "non-transport" ones.

The XCASCADE-3D predictions demonstrate how the electron transport affects the spatial distribution of excited electrons. For example, for the wide (micrometer) focusing, Figs.~\ref{fig:density0fs}a-b, the transport and non-transport electron distributions coincide with each other, suggesting that the electron ballistic transport could be neglected under these conditions. In contrast, in Fig.~\ref{fig:density0fs}f, the electron distributions along X axis and Y axis strongly differ from the respective non-transport  ones, indicating that the electron transport does play a significant role in case of tight focusing. The electron density along X spreads significantly wider than that along Y axis. This is expected, due to the x-ray polarization directed along X axis. Similar behaviour can be observed in Figs.~\ref{fig:density0fs}c-e at (still) submicrometer focusing.

The observations can be explained in the following way.  As we mentioned earlier, the electron range under the considered conditions is about 312 nm. The larger the beam spot is, the lower is the effect of the ballistic electron transport on the spatial redistribution of the absorbed X-ray energy. For the beam spots much larger than the electron range, the respective effect of electron transport becomes negligible.

\begin{figure}[h!] 
	
	\centering 
	
	\includegraphics[width=0.45\textwidth]{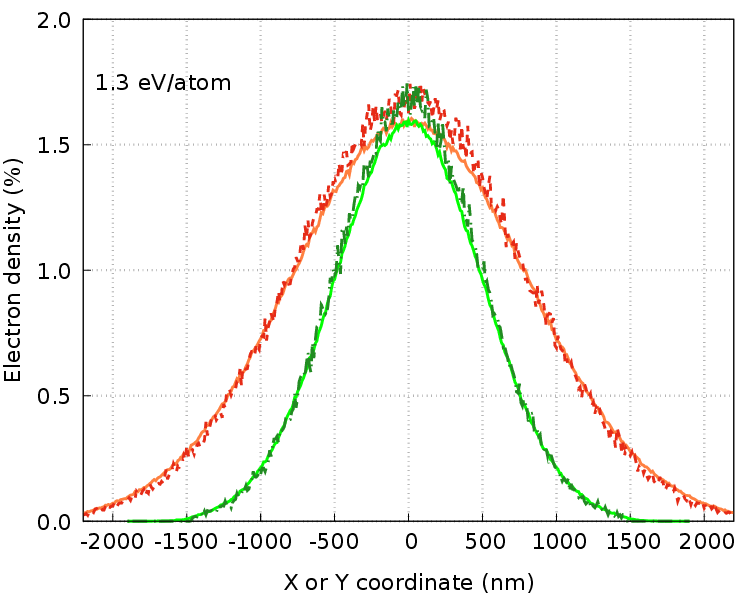}
	\includegraphics[width=0.45\textwidth]{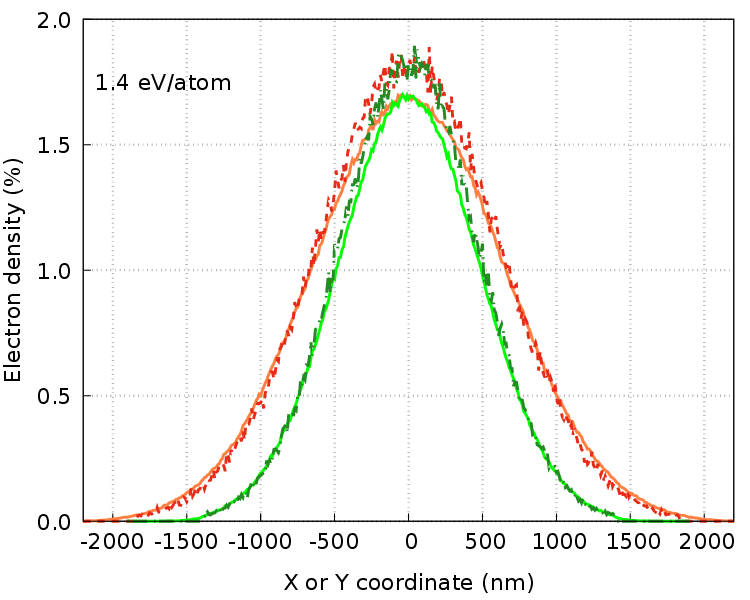}
	\includegraphics[width=0.45\textwidth]{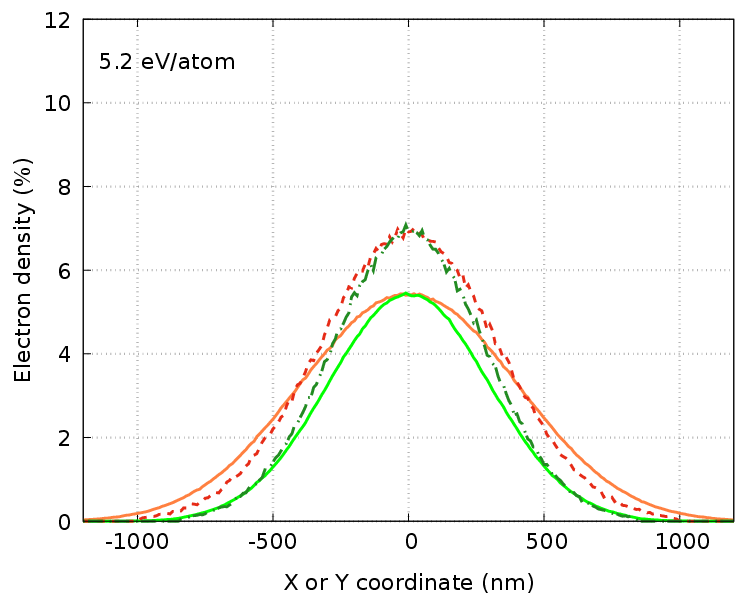} 
	\includegraphics[width=0.45\textwidth]{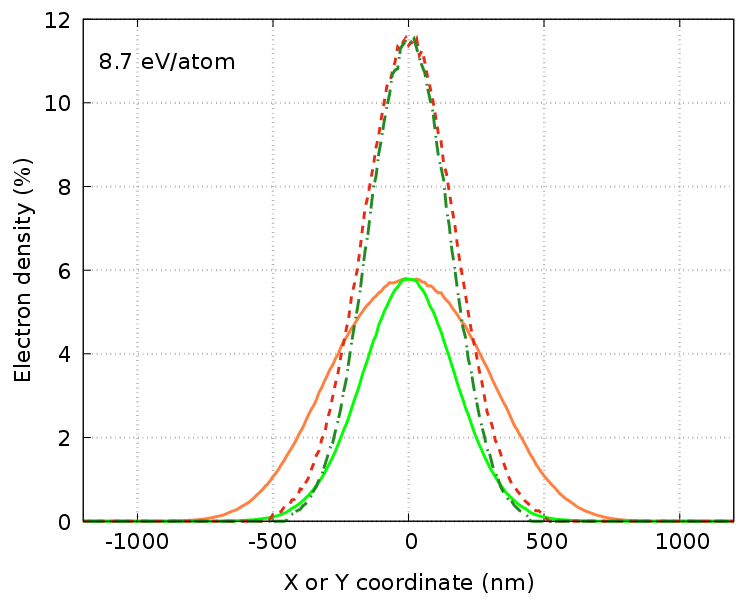} 
	\includegraphics[width=0.45\textwidth]{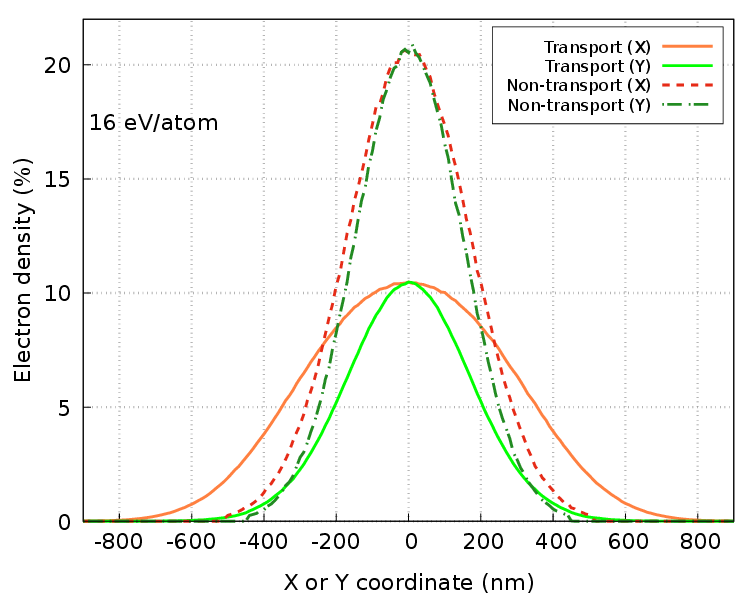} 
	\includegraphics[width=0.45\textwidth]{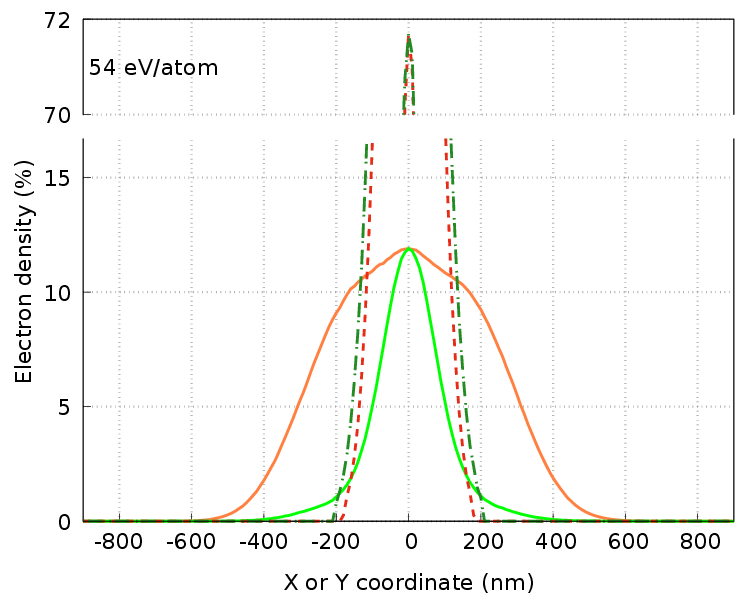} 
	
	\caption{Calculated electron density at time $t=30$~fs (i.e., after the cascading has finished). The other parameters and observables are the same as in Fig. \ref{fig:density0fs}.} 
	
	\label{fig:density30fs} 

\end{figure}

Fig.~\ref{fig:density30fs} shows the electron density distributions at 30 fs, when electron cascading was finished. As we mentioned earlier, the duration of the latter is about 14 fs. Taking into account the pulse duration of 6 fs, one can expect no ballistic transport after 20-30 fs. As in  Fig.~\ref{fig:density0fs}, dashed (dot-dashed) lines represent the simulated distribution of the absorbed photons multiplied by the average number of excited electrons per a single absorbed photon, which is about 422 at $t=30$ fs. In the present model, we do not take into account the electron diffusion for electrons with energies below the lowest ionization threshold (11.26 eV for diamond), i.e., we assume that such electrons completely stop their motion (see the justification in the previous section).

Similarly as in Fig.~\ref{fig:density0fs}, in Fig.~\ref{fig:density30fs} one can see the increasing effect of the ballistic electron transport on the spatial redistribution of absorbed energy when decreasing the beam spot size. The curves show that for tight focusing conditions (with the focus size smaller or comparable to the electron range), the electron spread caused by ballistic electrons significantly reduces the peak electron density, affecting also wing regions.

The resulting effective distribution of the absorbed X-ray dose determines the probability of diamond graphitization.


\section*{Discussion}

\label{sec:discussion}

According to our earlier simulations with two hybrid codes XTANT (relying on transferable tight binding description of solid's band structure) \cite{medvedev4open,Tavella2017graphitization} and XTANT+ (relying on density functional tight binding of the band structure) \cite{lipp2022density}, diamond undergoes ultrafast graphitization when the number of excited electrons in the conduction band reaches about 2.14\% or 4.60\%, respectively. This value corresponds to the ratio of the number of excited electrons to the initial number of valence electrons, which is 4 per atom in diamond. Note that the XCASCADE-3D model applied in this work is based on the atomistic approximation and does not include the electron-hole recombination or equilibration \cite{lipp2022quantifying}. However, these effects should not play a significant role on the considered  ultrafast timescales, up to 30 fs.

The simulation results presented in the previous section show that the considered experimental conditions lead to the peak electron densities presented in Table \ref{tab:irradiation_parameters}, last column. One can conclude that the conditions achieved in cases (3.1) and (3.2) correspond to the case of ultrafast amorphization \cite{lipp2022density}, with the peak dose too high for the graphitization to occur. According to our XTANT+ calculations \cite{lipp2022density}, the amorphization threshold for electron density lies between 12\% to 16\%. In contrast, in cases (1.1)-(1.4) the peak electron density is too low to expect any graphitization.

In cases (2.1)-(2.3), the predictions of the two hybrid codes differ. XTANT predictions indicate that even in these cases the electron density were too high to achieve the graphitization, while XTANT+ predictions suggest that the partial graphitization could be possible in the near-peak region on the timescale of about 500 fs. In Ref. \cite{Heimann2023Nonthermal}, the signal was too weak to observe such partial graphitization on the experimentally accessible ultrafast time scale (up to 300 fs).

Cases (3.1), (3.2) or (2.1)-(2.3) might result in graphitization regions located in the wing region of the pulse, and further apart, although a possible heat transfer from the peak to the wings could hinder the graphitization process.

One of the limitations of the XCASCADE-3D model is that it relies on the assumption of the independent electron cascades. This assumption is valid when the density of cascading electrons is considerably lower than the atomic density, i.e., at least ten times smaller than the atomic density. The condition also justifies the usage of cross sections for unexcited materials in the code. 
We have checked the validity of the above condition for the cases presented in Figs.~\ref{fig:density0fs} and \ref{fig:density30fs}, i.e., when cascading process was still on-going or when it finished, respectively. The plots represent the normalized density of excited electrons (the ratio of the number of excited electrons to the number of valence electrons which is here 4 per atom). Therefore, the limiting electron density value is 2.5\%. This is the case for all plots in Fig.~\ref{fig:density0fs} and Fig.~\ref{fig:density30fs}a,b. However in Figs.~\ref{fig:density30fs}c-f, the electron density increases above this limit, i.e., the XCASCADE-3D predictions are no longer reliable and should be treated only as indicative. In any case, the obtained prediction span confirms a need of the careful consideration of each specific dose case at the stage of experimental planning.

In summary, triggering graphitization or other phase transitions by hard X-ray photons is difficult due to the escape of electrons from the beam focus. Decisive parameter here is the electron range, i.e., how it compares to the focal beam size. In hard x-ray regime discussed above, the released photo- and Auger electrons can be very energetic, e.g., for 7 keV photons the photoelectron range amounts to 312 nm. In case of submicrometer focusing, such electrons can strongly influence redistribution of the X-ray absorbed dose deposited in various lateral regions of the crystal. In contrast, in case of soft x-ray induced ultrafast graphitization, indirectly observed in \cite{Tavella2017graphitization}, the electron range was only $\sim 0.8$ nm vs. micrometer beam focus, i.e., the absorbed X-ray energy was not carried out by the electrons out of the beam focus in this case.

Therefore, for future experiments on x-ray induced phase transitions, dedicated simulations are necessary for establishing optimal experimental conditions ensuring a required dose distribution.

\begin{acknowledgments}
We thank DESY Maxwell cluster for computational resources provided.

\end{acknowledgments}

\bibliography{lipp-own,mybib}

\end{document}